\documentclass[journal]{IEEEtran}

\usepackage{graphicx}
\usepackage{grffile}
\usepackage{amssymb}
\usepackage{amsmath}
\usepackage{amsthm}
\usepackage{algorithm}
\usepackage{algorithmic}
\usepackage[usenames,dvipsnames]{color}
\usepackage{multicols}

\begin{document}
\title{Pattern Clustering using Cooperative Game Theory}

\author{Swapnil~Dhamal,
        Satyanath~Bhat,
        Anoop~K~R,
        and~Varun~R~Embar}%

\markboth{Centenary Conference, 2011 - Electrical Engineering, Indian Institute of Science, Bangalore}%
{Shell \MakeLowercase{\textit{et al.}}: Pattern Clustering using Cooperative Game Theory}
\maketitle

\begin{abstract}

In this paper, we approach the classical problem of clustering using solution concepts from cooperative game theory such as Nucleolus and Shapley value. We formulate the problem of clustering as a characteristic form game and develop a novel algorithm DRAC (Density-Restricted Agglomerative Clustering) for clustering. With extensive experimentation on standard data sets, we compare the performance of DRAC with that of well known algorithms. We show an interesting result that four 
prominent solution concepts, Nucleolus, Shapley value, Gately point and $\tau$-value coincide for the defined characteristic form game. This vindicates the choice of the characteristic function of the clustering game and also provides strong intuitive foundation for our approach.
\end{abstract}

\begin{IEEEkeywords}
Pattern clustering, Characteristic form game, Nucleolus, Shapley value.
\end{IEEEkeywords}

\section{Introduction}
\label{sec:intro}
\IEEEPARstart{C}{lustering} or unsupervised classification of patterns into groups based on similarity is a very well studied problem in pattern recognition, data mining, information retrieval, and related disciplines. Besides, clustering has also been used in solving extremely large scale problems. Clustering also acts as a precursor to many data processing tasks including classification. According to Backer and Jain~\cite{backerjain}, “in cluster analysis, a group of objects is split into a number of more or less homogeneous subgroups on the basis of an often subjectively chosen measure of similarity (i.e., chosen subjectively based on its ability to create “interesting” clusters) such that the similarity between objects within a subgroup is larger than the similarity between objects belonging to different subgroups“.
A key problem in the clustering domain is to determine the number of output clusters $k$. Use of cooperative game theory provides a novel way of addressing this problem by using a variety of solution concepts.

In the rest of this section, we justify the use of game theoretic solution concepts, specifically Nucleolus, for pattern clustering, give an intuition why the various solution concepts coincide and refer to a few recent works in clustering using game theory. In Section~\ref{sec:prelim}, we provide a brief introduction to the relevant solution concepts in cooperative game theory. Sections~\ref{sec:gamemodel} explains our model and algorithm for clustering based on cooperative game theory. In Section~\ref{sec:results}, we describe the experimental results and provide a comparison of our algorithm with some existing related ones. The coincidence of Nucleolus, Shapley value, Gately point and $\tau$-value with the chosen characteristic function is discussed and formally proved in Section~\ref{sec:equiv_nuc_shap}. We conclude with future work in Section~\ref{sec:conclusion}.

We motivate the use of game theory for pattern clustering  with an overview of a previous approach. SHARPC~\cite{sharpc} proposes a novel approach to find the cluster centers in order to give a good start to K-means, which thus results in the desired clustering. The limitation of this approach is that it is restricted to K-means, which is not always desirable especially when the classes have unequal variances or when they lack convex nature. 
We, therefore, extend this approach to a more general clustering problem in ${\mathbb{R}}^2$.

As it will be clear in Section~\ref{sec:prelim}, Shapley value is based on average fairness, Gately point is based on stability, $\tau$-value is based on efficiency while Nucleolus is based on both min-max fairness and stability. Hence, it is worthwhile exploring these solution concepts to harness their properties for the clustering game. Of these solution concepts, the properties of Nucleolus, viz., fairness and stability, are the most suitable for the clustering game. Moreover, we show in Section~\ref{sec:equiv_nuc_shap} that all these solution concepts coincide for the chosen characteristic function. As finding Nucleolus, for instance, is computationally expensive, it is to our advantage if we use the computational ease of other solution concepts. We see in Section~\ref{sec:gamemodel} that for the chosen characteristic function, the Shapley value can be computed in polynomial time. So for our algorithm, we use Shapley value, which is equivalent to using any or all of these solution concepts.

The prime reason for the coincidence of the relevant solution concepts is that the core, which we will see in Section~\ref{core}, is symmetric about a single point and all these solution concepts coincide with that very point. We will discuss this situation in detail and prove it formally in Section~\ref{sec:equiv_nuc_shap}.

There have been approaches proposing the use of game theory for pattern clustering. Garg, Narahari and Murthy \cite{sharpc} propose the use of Shapley value to give a good start to K-means. 
Gupta and Ranganathan \cite{upavan1,upavan2} use a microeconomic game theoretic approach for clustering, which simultaneously optimizes two objectives, viz. compaction and equipartitioning. Bulo and Pelillo \cite{evolution} use the concept of evolutionary games for hypergraph clustering. Chun and Hokari \cite{neqs} prove the coincidence of Nucleolus and Shapley value for queueing problems.

The contributions of our work are as follows:
\begin{itemize}
\item We explore game theoretic solution concepts for the clustering problem.
\item We prove coincidence of Nucleolus, Shapley value, Gately point and $\tau$-value for the defined game.
 \item We propose an algorithm, DRAC (Density-Restricted Agglomerative Clustering), which overcomes the limitations of K-means, Agglomerative clustering, DBSCAN \cite{dbscanpaper} and OPTICS \cite{opticspaper} using game theoretic solution concepts.
\end{itemize}
\section{Preliminaries}
\label{sec:prelim}
In this section, we provide a brief insight into the cooperative game theory concepts~\cite{straffin,coaltut,neqs} viz. Core, Nucleolus, Shapley value, Gately point and $\tau$-value.

A cooperative game $(N,\nu)$ consists of two parameters $N$ and $\nu$. $N$ is the set of players and $\nu:2^N\rightarrow \mathbb{R}$ is the characteristic function. It defines the value $\nu(S)$ of any coalition $S\subseteq N$. 
\subsection{The Core}
\label{core}
Let $(N,\nu)$ be a coalitional game with transferable utility (TU). Let $x = (x_1,\ldots,x_n)$, where $x_i$ represents the payoff of player $i$, the core consists of all payoff allocations $x = (x_1, . . . , x_n)$ that satisfy the following properties.
\begin{enumerate}
 \item individual rationality, i.e., $x_i \geq \nu(\{i\}) \; \forall \; i \in N$
 \item collective rationality i.e. $ \sum_{i \in N} x_i = \nu(N)$.
 \item coalitional rationality i.e. $ \sum_{i \in S} x_i \geq \nu(S) \; \forall S \subseteq N$.
\end{enumerate}
A payoff allocation satisfying individual rationality and collective rationality is called an \textit{imputation}. 
\subsection{The Nucleolus}

Nucleolus is an allocation that minimizes the dissatisfaction of the players from the allocation they can receive in a game~\cite{nucleolus}. For every imputation $x$, consider the excess defined by
\begin{displaymath}
 e_S(x) = \nu(S) - \sum_{i \in S} x_i
\end{displaymath}
$e_S(x)$ is a measure of unhappiness of $S$ with $x$. The goal of Nucleolus is to minimize the most unhappy coalition, i.e., largest of the $e_S(x)$. 
The linear programming problem formulation is as follows
\begin{displaymath}
\text{min }Z 
\end{displaymath}
subject to
\begin{displaymath}
 Z + \sum_{i \in S} x_i \geq \nu(S) \; \: \forall S \subseteq N
\end{displaymath}
\begin{displaymath}
\sum_{i \in N} x_i = \nu(N)
\end{displaymath}
The reader is referred to \cite{coaltut} for the detailed properties of Nucleolus. It combines a number of fairness criteria with stability. It is the imputation which is lexicographically central and thus fair and optimum in the min-max sense.

\subsection{The Shapley Value}
\label{sec:shapley}

Any imputation $\phi = (\phi_1, . . . ,\phi_n)$ is a Shapley value if it follows the axioms which are based on the idea of fairness. The reader is referred to~\cite{straffin} for the detailed axioms.
For any general coalitional game with transferable utility $(N,\nu)$, the Shapley value of player $i$ is given by

\begin{eqnarray*}
\nonumber
 \phi_i &=& \frac{1}{n!} \sum_{i\in S} (|S|-1)!(n-|S|)![\nu(S)-\nu(S-i)] \\
 &=&\frac{1}{n!}\sum_{\pi\in \Pi}x_i^\pi
\end{eqnarray*}
$\Pi=$ set of all permutations on N\\
$x_i^\pi=$ contribution of player $i$ to permutation $\pi$

\subsection{The Gately Point}
\label{sec:gately}
Player $i$'s \textit{propensity to disrupt} the grand coalition is defined to be the following ratio~\cite{straffin}.
\begin{equation}
\label{gatelyeq}
d_i(x) = \frac{\sum_{j\neq i}x_j - \nu(N-i)}{x_i - \nu(i)}
\end{equation}
If $d_i(x)$ is large, player $i$ may lose something by deserting the grand coalition, but others will lose a lot more. The Gately point of a game is the imputation which minimizes the maximum propensity to disrupt. The general way to minimize the largest propensity to disrupt is to make all of the propensities to disrupt equal. When the game is normalized so that $\nu(i) = 0$ for all $i$, the way to set all the $d_i(x)$ equal is to choose $x_i$ in proportion to $\nu(N) - \nu(N-i)$.
\begin{equation}
\nonumber
Gv_i = \frac{\nu(N) - \nu(N-i)}{\sum_{j\in N}(\nu(N) - \nu(N-j))}\nu(N)
\end{equation}

\subsection{The $\tau$-value}
$\tau$-value is the unique solution concept which is efficient and has the \textit{minimal right property} and the \textit{restricted proportionality property}. 
The reader is referred to \cite{tau} for the details of these properties.
For each $i\in N$, let
\begin{equation}
\label{taumi}
M_i(\nu) = \nu(N) - \nu(N-i) \text{ and }m_i(\nu) = \nu(i)
\end{equation} 
Then the $\tau$-value selects the maximal feasible allocation on the line connecting $M(\nu) = (M_i(\nu))_{i\in N}$ and $m(\nu) = (m_i(\nu))_{i\in N}$~\cite{neqs}.
For each convex game $(N,\nu)$,
\begin{equation}
\label{taueq}
\tau(\nu) = \lambda M(\nu) + (1-\lambda)m(\nu)
\end{equation}
where $\lambda \in [0,1]$ is chosen so as to satisfy
\begin{equation}
\label{tausatis}
\sum_{i\in N} [\lambda(\nu(N)-\nu(N-i)) + (1-\lambda)\nu(i)] = \nu(N)
\end{equation}

\section{A Model and Algorithm for Clustering based on Cooperative Game Theory}\label{sec:gamemodel}
For the clustering game, the characteristic function is chosen as in~\cite{sharpc}.
\begin{equation}
 \nu(S)=\frac{1}{2}\sum_{\substack{i, j \in S \\i \neq j}} f(d(i,j))
\label{charfn}
\end{equation}
In Equation~\ref{charfn}, $d$ is the \textit{Euclidean distance}, $f:d\rightarrow [0,1]$ is a similarity function. Intuitively, if two points $i$ and $j$ have small euclidean distance, then $f(d(i,j))$ approaches 1. 
The similarity function that we use in our implementation is
\begin{equation}
f(d(i,j)) = 1-\frac{d(i,j)}{d_{M}}
 \label{similarityfn}
\end{equation}
where $d_{M}$ is the maximum of the distances between all pairs of points in the dataset.

When Equation~\ref{charfn} is used as characteristic function, it is shown in \cite{sharpc} that Shapley value of player $i$ can be computed in polynomial time and is given by
\begin{equation}
 \phi_i= \frac{1}{2} \sum_{\substack{j \in N\\j \neq i}}f(d(i,j))
 \label{shapleyeq}
\end{equation}

Also, from Equation \ref{charfn}, it can be derived that
\begin{equation}
\nu(S) = \sum_{\substack{T\subseteq S \\ |T|=2}}\nu(T)
 \label{pair}
\end{equation}

In Sections~\ref{sec:intro} and \ref{sec:prelim}, we have discussed the benefits of imputations resulting from various game theoretic solution concepts. Also, in Section~\ref{sec:equiv_nuc_shap}, we will show that all these imputations coincide. Moreover, as Equation \ref{shapleyeq} shows the ease of computation of Shapley value in the clustering game with the chosen characteristic function, we use Shapley value as the base solution concept for our algorithm.

The basic idea behind the algorithm is that we expand the clusters based on density. From Equations \ref{similarityfn} and \ref{shapleyeq}, Shapley value represents density in some sense. For every cluster, we start with an unallocated point with the maximum Shapley value and assign it as the cluster center. If that point has high density around it, it should only consider the close-by points, otherwise it should consider more faraway points. We implement this idea in step \ref{calcbeta} of Algorithm \ref{dracalgo} with parameter $\beta$. For the point with the globally maximum Shapley value, $\beta=\delta$, while it is low for other cluster centers. Also, as we go from cluster center with the highest Shapley value to those with lower values, we do not want to degrade the value of $\beta$ linearly. So we have square-root function in step \ref{calcbeta}. Alternatively, it can be replaced with any other function which ensures sub-linear degradation of $\beta$. The input parameters $\delta$ and $\gamma$ should be changed accordingly.

 \begin{algorithm}[H]

\caption{Density-Restricted Agglomerative Clustering (DRAC)}
\begin{algorithmic}[1]
\label{dracalgo}
\REQUIRE {Dataset, maximum threshold for similarity $\delta \in [0,1]$  and threshold for Shapley value multiplicity $\gamma \in [0,1]$}
\STATE Find the pairwise similarity between all points in dataset.
\STATE For each point $i$, compute the Shapley value using Equations \ref{similarityfn} and \ref{shapleyeq}.
\STATE Arrange the points in non-increasing order of their Shapley values. Let $g_M$ be the global maximum of Shapley values. Start a new queue, let's call it \textit{expansion queue}.
\STATE \label{startagain} Start a new cluster. Of all the unallocated points, choose the point with maximum Shapley value as the new cluster center. Let $l_M$ be its Shapley value. Mark that point as allocated. Add it to the expansion queue.
\STATE \label{calcbeta} Set $\beta=\delta \sqrt{\frac{l_M}{g_M}}$.
\STATE \label{substart} For each unallocated point, if the similarity of that point to the first point in the expansion queue is at least $\beta$, add it to the current cluster and mark it as allocated. 
If the Shapley value of that point is at least $\gamma$-multiple of $l_M$, add it to the expansion queue.
\STATE Remove the first point from the expansion queue.
\STATE If the expansion queue is not empty, go to step \ref{substart}.
\STATE If the cluster center is the only point in its cluster, mark it as noise.
\STATE If all points are allocated a cluster, terminate. Else go to step \ref{startagain}.
\end{algorithmic}
\end{algorithm}

Secondly, when the density around a point is very low as compared to the density around the cluster center of the cluster of which it is a part of, it should not be responsible for further growth of the cluster. This ensures that clusters are not merged together when they are connected with a thin bridge of points. It also ensures that the density within a cluster does not vary beyond a certain limit. We implement this idea with what we call an expansion queue. We add points to the queue only if their Shapley value is at least $\gamma$-multiple of that of the cluster center of the cluster of which it is a part of. The expansion queue is responsible for the growth of a cluster and it ceases once the queue is empty. The detailed and systematic steps are given in Algorithm \ref{dracalgo}.

\section{Experimental Results}
\label{sec:results}
In this section, we qualitatively compare our algorithm with some existing related algorithms. SHARPC \cite{sharpc} gives a good start to K-means using a game theoretic solution concept, viz., the Shapley value. As our algorithm hierarchically allocates points to the cluster starting from a cluster center, we compare it with Agglomerative Clustering. The way our characteristic function and similarity function are defined, the Shapley value represents density in some sense. So we compare our algorithm with the density-based ones, viz., DBSCAN (Density-Based Spatial Clustering of Applications with Noise) and OPTICS (Ordering Points To Identify the Clustering Structure). 
Throughout this section, `cluster ($<$\textit{colored marker}$>$)' refers to the cluster marked by that colored marker in the corresponding figure. Noise is represented by (\textcolor{black}{$\circ$}).

 \begin{figure}[h]
\begin{center}
    \includegraphics[scale=.6]{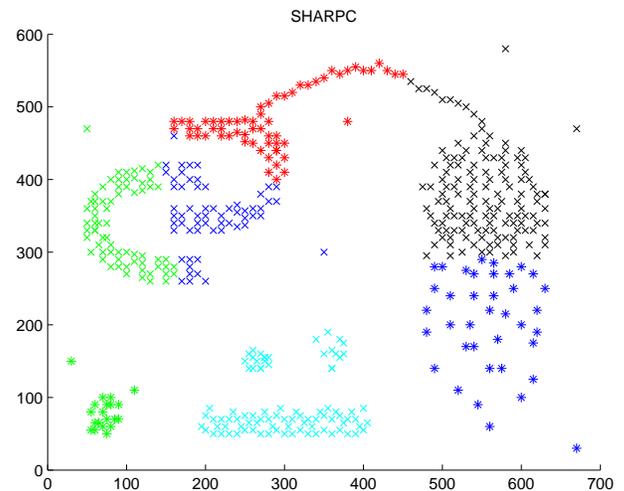}
    \caption{Clusters as discovered by SHARPC}
    \label{sharpcfig}
\end{center}
   \end{figure}

 \begin{figure}[h]
\begin{center}
    \includegraphics[scale=.6]{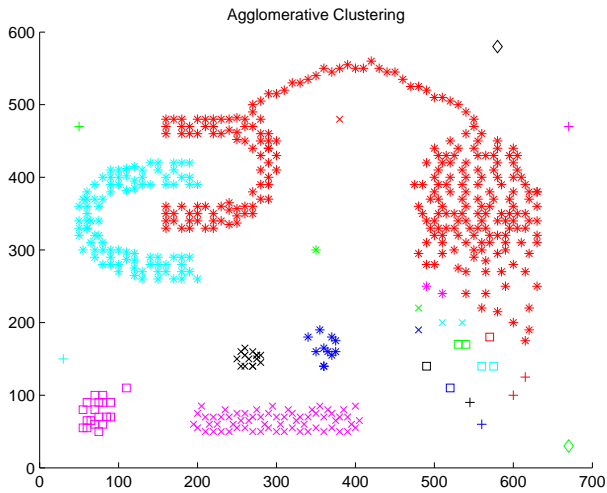}
    \caption{Clusters as discovered by Agglomerative Clustering}
    \label{acfig}
\end{center}
   \end{figure}

 \begin{figure}[h]
\begin{center}
    \includegraphics[scale=.6]{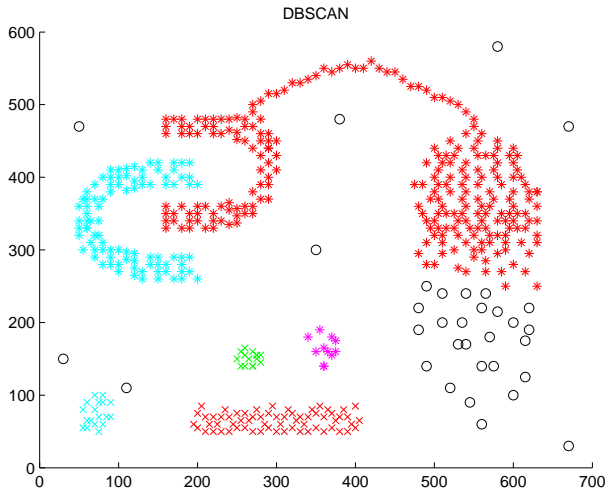}
    \caption{Clusters as discovered by DBSCAN}
    \label{dbscanfig}
\end{center}
   \end{figure}

 \begin{figure}[h]
\begin{center}
    \includegraphics[scale=.6]{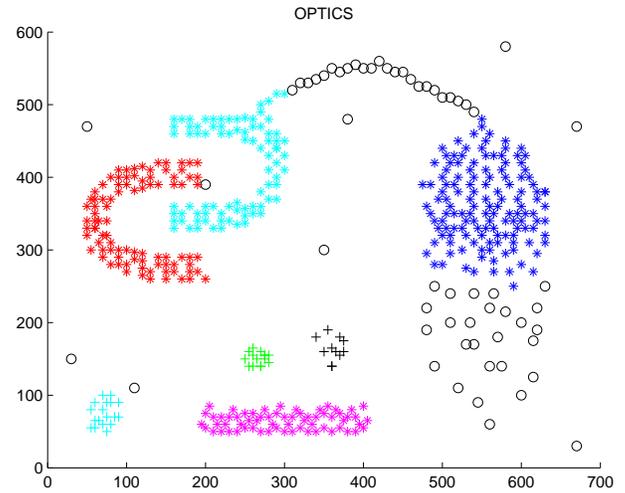}
    \caption{Clusters as discovered by OPTICS}
    \label{opticsfig}
\end{center}
   \end{figure}

 \begin{figure}[h]
\begin{center}
    \includegraphics[scale=.6]{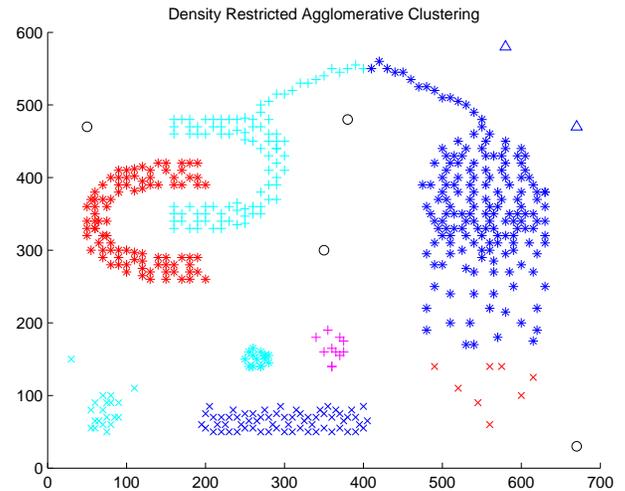}
    \caption{Clusters as discovered by DRAC}
    \label{dracfig}
\end{center}
   \end{figure}

Figure \ref{sharpcfig} shows the clusters formed by SHARPC \cite{sharpc} which tries to allocate clusters by enclosing points in equal-sized spheres. It cannot detect clusters that are not convex. Also, the cluster (\textcolor{cyan}{$\times$}) is a merging of three different clusters. If the threshold is increased so as to solve the second problem, more clusters are formed and the larger clusters get subdivided into several smaller clusters.

Agglomerative Clustering, as Figure \ref{acfig} shows, can detect clusters of any shape and size. But owing to a constant threshold for the growth of all clusters, it faces the problem of forming several clusters in the lower right part when they should have been part of one single cluster. If the threshold is decreased so as to solve this problem, clusters (\textcolor{red}{$*$}) and (\textcolor{cyan}{$*$}) get merged. Another problem is that the bridge connecting the two classes merges these into one single cluster (\textcolor{red}{$*$}).

Figure \ref{dbscanfig} shows the results of DBSCAN \cite{dbscanpaper}. It is well known that it cannot detect clusters with different densities in general. The points in the lower right part are detected as noise when intuitively, the region is dense enough to be classified as a cluster. An attempt to do so compromises the classification of clusters (\textcolor{red}{$*$}) and (\textcolor{cyan}{$*$}) as distinct. Moreover, the bridge connecting the two classes merges them into one single cluster (\textcolor{red}{$*$}). An attempt to do the required classification leads to unnecessary subdivision of the rightmost class and more points being detected as noise.

The clustering obtained using OPTICS \cite{opticspaper} is shown in Figure \ref{opticsfig}. Unlike DBSCAN, clusters (\textcolor{cyan}{$*$}) and (\textcolor{blue}{$*$}) are detected as distinct. However, the points in the lower right part are detected as noise when they should have been classified as one cluster. The reachability plots for different values of \textit{minpts} are such that an attempt to classify some of these points as a part of some cluster leads to the merging of clusters (\textcolor{red}{$*$}) and (\textcolor{cyan}{$*$}). If we continue trying to get more of these points allocated, the bridge plays the role of merging the two clusters (\textcolor{cyan}{$*$}) and (\textcolor{blue}{$*$}).

Figure \ref{dracfig} shows the clustering obtained using Density-Restricted Agglomerative Clustering (DRAC). As cluster (\textcolor{cyan}{$+$}) is highly dense, its cluster center has very high Shapley value resulting in a very high value of $\beta$, the similarity threshold. No point in cluster (\textcolor{red}{$*$}) crosses the required similarity threshold with the points in cluster (\textcolor{cyan}{$+$}), thus ensuring that the two clusters are not merged. The points in the central part of the bridge have extremely low Shapley values as compared to the cluster center of cluster (\textcolor{cyan}{$+$}) and so they fail to cross the Shapley value threshold of having at least $\gamma$-multiple of the Shapley value of the cluster center. This ensures that they are not added to the expansion queue of the cluster, thus avoiding the cluster growth to extend to cluster (\textcolor{blue}{$*$}). Cluster (\textcolor{blue}{$*$}) extends to the relatively low density region because of points being added to the expansion queue owing to their sufficiently high Shapley value, at least $\gamma$-multiple of the Shapley value of the cluster center. Cluster (\textcolor{red}{$\times$}) is a low density cluster owing to the low Shapley value of the cluster center and so low value of $\beta$, the similarity threshold, thus allowing more faraway points to be a part of the cluster. Cluster centers, which fail to agglomerate at least one point with their respective values of $\beta$, are marked as noise.

Like other clustering algorithms, Algorithm \ref{dracalgo} faces some limitations. As it uses Equations \ref{similarityfn} and \ref{shapleyeq} to compute the Shapley values, the Shapley value of a point changes even when a remote point is altered, which may change its cluster allocation. For the same reason, the Shapley values of the points close to the mean of the whole dataset is higher than other points even when the density around them is not as high. One solution to this problem is to take the positioning of the point into account while computing its Shapley value. There is no explicit noise detection. A point is marked as noise if it is the only point in its cluster. For instance, in Figure \ref{dracfig}, the two points in the upper right corner are noise points, but owing to their low Shapley values, $\beta$ is very low and so they are classified as a separate cluster (\textcolor{blue}{$\bigtriangleup$}) instead. The amortized time complexity of Algorithm \ref{dracalgo} is $O(n^2)$.

\newcommand{\BlackBox}{\rule{1.5ex}{1.5ex}}  
\newenvironment{proofblackdot}{\par\noindent{\bf Proof\
}}{\hfill\BlackBox\\[2mm]}
\newtheorem{example}{Example}
\newtheorem{theorem}{Theorem}
\newtheorem{lemma}{Lemma}
\newtheorem{proposition}{Proposition}
\newtheorem{claim}[]{Claim}
\newtheorem{remark}[theorem]{Remark}
\newtheorem{corollary}[theorem]{Corollary}
\newtheorem{definition}[theorem]{Definition}
\newtheorem{conjecture}[theorem]{Conjecture}
\newtheorem{axiom}[theorem]{Axiom}
\section{Coincidence of Nucleolus, Shapley value, Gately point and $\tau$-value in the current setting}
\label{sec:equiv_nuc_shap}

In the game as defined in Section \ref{sec:gamemodel}, we show in this section, that Nucleolus, Shapley value, Gately point and $\tau$-value coincide. First, we discuss the structure of the core. The core is symmetric about a single point, which is the prime reason why the above solution concepts coincide with that very point.

 \begin{figure}[h]
\begin{center}
    \includegraphics[scale=.17]{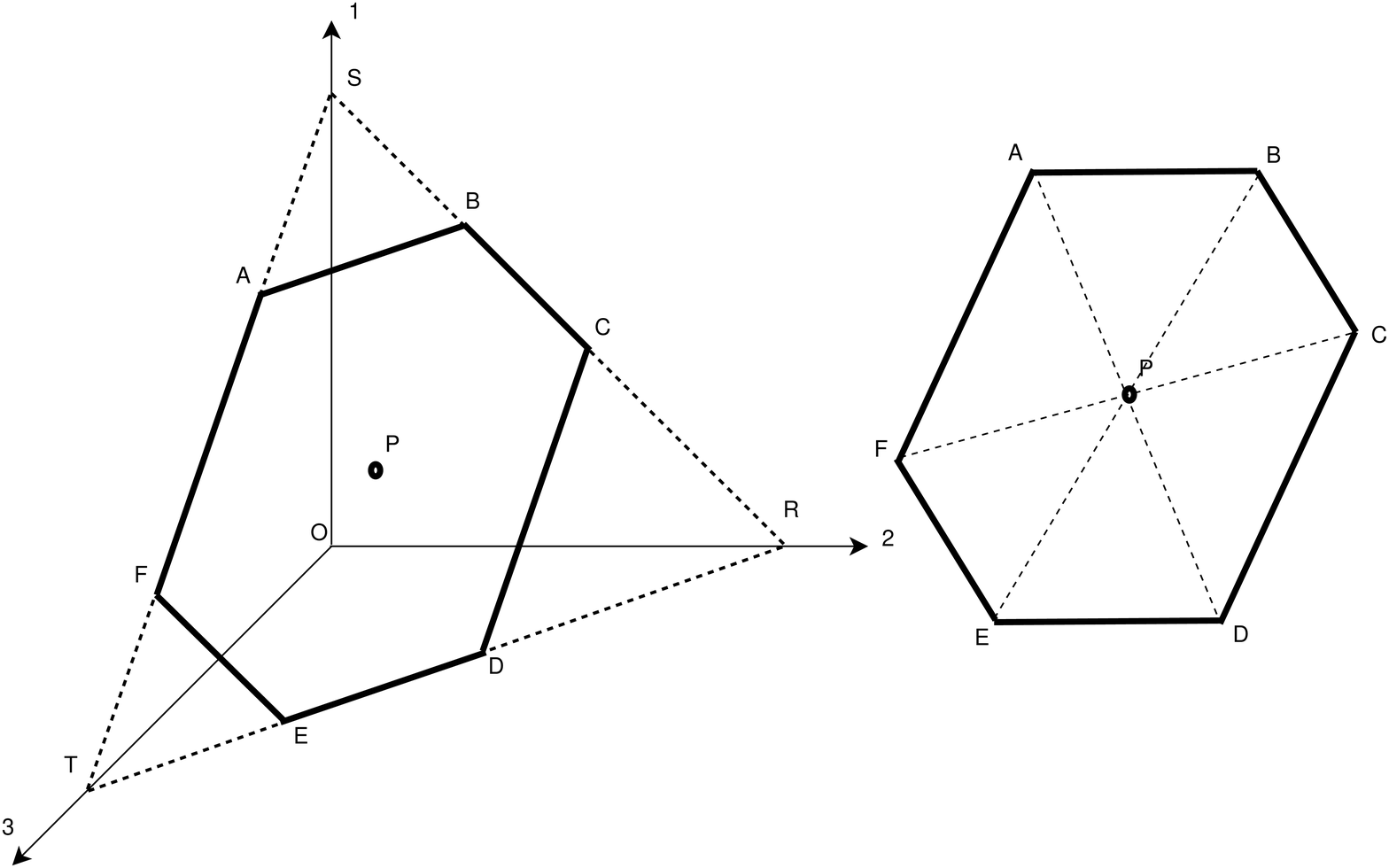}
    \caption{The game has a symmetric core. This figure shows the core for a 3-player game.}
    \label{symm}
\end{center}
   \end{figure}

Figure~\ref{symm} shows the core for a 3-player cooperative game, in our case, a 3-point clustering game. The $STR$ plane corresponds to collective rationality constraint, sides $AF, BC, DE$ correspond to individual rationality constraints while sides $AB, CD, EF$ correspond to coalitional rationality constraints. The reader is referred to \cite{straffin} for a detailed discussion on imputation triangle of a 3-player cooperative game. By simple geometry and theory on imputation triangle, it can be seen that $AB = DE = \nu(\{2,3\})\sqrt{2}$. Similarly, all opposite sides of the core are equal and so the core is symmetric about its center $P$.

Clearly, any point other than $P$ will have more distance from at least one side and so will be lexicographically greater than $P$, which means that $P$ is the Nucleolus. Also, as the core is symmetric, it is intuitive that $P$ is the fairest of all allocations, which means that it corresponds to the Shapley value imputation. We prove a general result for $n$-player clustering game that all the relevant solution concepts coincide.

\begin{proposition}
\label{prop}
 For the transferable utility (TU) game defined by Equation \ref{charfn}, for each $i \in N$, the Shapley Value is given by
\begin{equation}
\label{shapleyeqhalf}
\phi_i = \frac{1}{2}\sum_{\substack{S \subseteq N\\ i \in S\\|S| = 2}} \nu(S)
\end{equation}
\end{proposition}

\begin{proofblackdot}
From Equations \ref{charfn} and \ref{shapleyeq},
\begin{eqnarray*}
\nonumber
\phi_i &=& \frac{1}{2} \sum_{\substack{j \in N\\j \neq i}}f(d(i,j)) \\
&=& \frac{1}{2}\ \frac{1}{2} \sum_{\substack{S \subseteq N\\|S|=2\\k, l \in S\\ k \neq l\\ i \in S}}f(d(k,l))\\
&=& \frac{1}{2}\sum_{\substack{S \subseteq N\\|S|=2\\\\ i \in S}} \frac{1}{2}\sum_{\substack{k, l \in S\\ k \neq l}}f(d(k,l)) \\
&=& \frac{1}{2}\sum_{\substack{S \subseteq N\\|S|=2\\\\ i \in S}} \nu(S)
\end{eqnarray*}
\end{proofblackdot}

\begin{lemma} \cite{neqs}
\label{lem1}
 For the TU game satisfying Equation \ref{shapleyeqhalf}, for each $S \subseteq N$,
\begin{equation}
 \nonumber
 \nu(S) - \sum_{i \in S}\phi_i = \nu(N\backslash S) - \sum_{i \in N\backslash S}\phi_i
\end{equation}
\end{lemma} 
\begin{flushleft}
The reader is referred to \cite{neqs} for the proof of Lemma~\ref{lem1}.
\end{flushleft}

\begin{theorem} \cite{neqs}
\label{shapeqnu}
 For the TU game satisfying Equation~\ref{shapleyeqhalf},
\begin{equation}
\nonumber
 \phi(\nu)=Nu(\nu)
\end{equation}
where $Nu(\nu)$ is the Nucleolus of the TU game $(N,\nu)$.
\end{theorem}
\begin{flushleft}
 The reader is referred to \cite{neqs} for the proof of Theorem~\ref{shapeqnu}.
\end{flushleft}

\begin{theorem}

\label{shapeqgate}
 For the TU game defined by Equation \ref{charfn},
\begin{equation}
\nonumber
 \phi(\nu)=Gv(\nu)
\end{equation}
where $Gv(\nu)$ is the Gately point of the TU game $(N,\nu)$.

\end{theorem}

\begin{proofblackdot}
By Lemma \ref{lem1}, when $S = \{i\}$, we have
\begin{equation}
\nonumber
 \nu(i) - \phi_i = \nu(N-i) - \sum_{j\neq i}\phi_j
\end{equation}
From Equation \ref{gatelyeq}, the propensity to disrupt for player $i$ when imputation is the Shapley value is
\begin{equation}
\nonumber
d_i(\phi) = \frac{\sum_{j\neq i}\phi_j - \nu(N-i)}{\phi_i - \nu(i)} = 1
\end{equation}
As the propensity to disrupt is 1 for every player $i$, it is equal for all the players and hence, from the theory in Section~\ref{sec:gately}, the Shapley value imputation is the Gately point.
 \begin{equation}
 \nonumber
  \phi(\nu)=Gv(\nu)
 \end{equation}
\end{proofblackdot}

\begin{theorem}
\label{shapeqtau}
 For the TU game defined by Equation \ref{charfn},
\begin{equation}
\nonumber
 \phi(\nu)=\tau(\nu)
\end{equation}
where $\tau(\nu)$ is the $\tau$-value of the TU game $(N,\nu)$.

\end{theorem}

\begin{proofblackdot}
From Equations \ref{taumi} and \ref{pair},
\begin{eqnarray*}
M_i(\nu)
&=& \nu(N)-\nu(N-i) \\
&=& \sum_{\substack{S\subseteq N \\ |S| = 2}}\nu(S) - \sum_{\substack{S\subseteq N\backslash \{i\} \\ |S| = 2}}\nu(S) \\
&=& \sum_{\substack{S\subseteq N \\ i \in S \\ |S| = 2}}\nu(S)
\end{eqnarray*}
This, with Equation \ref{tausatis} and the fact that for our $(N,\nu)$ game, for all $i$, $m_i(\nu) = \nu(i) = 0$,
\begin{eqnarray*}
 \nu(N) &=& \lambda\sum_{i\in N}M_i(\nu) \\ 
&=& \lambda\sum_{i\in N} \sum_{\substack{S\subseteq N \\ i \in S \\ |S| = 2}}\nu(S)\\
&=& 2\lambda\sum_{\substack{S\subseteq N \\ |S| = 2}}\nu(S) \\ 
\end{eqnarray*}
Using Equation \ref{pair}, we get $\lambda = \frac{1}{2}$.
This, with Equation~\ref{taueq} and the fact that for all $i$, $m_i(\nu) = 0$,
\begin{equation}
 \nonumber
\tau_i(\nu) = \frac{1}{2}\sum_{\substack{S \subseteq N\\ i \in S\\|S| = 2}} \nu(S)
\end{equation}
This, with Proposition~\ref{prop}, gives 
 \begin{equation}
 \nonumber
 \phi(\nu) = \tau(\nu)
 \end{equation}
\end{proofblackdot}

From Theorem~\ref{shapeqnu}, Theorem~\ref{shapeqgate} and Theorem~\ref{shapeqtau}, the Nucleolus, the Shapley value, the Gately point and the $\tau$-value coincide in the clustering game with the chosen characteristic function. These results further vindicate our choice of characteristic function for the clustering game.

\section{Conclusion and Future Work} \label{sec:conclusion}
We have explored game theoretic solution concepts as an alternative to the existing methods, for the clustering problem. Also, Nucleolus being both min-max fair and stable, is the most suitable solution concept for pattern clustering. We have also proved the coincidence of Nucleolus, Shapley value, Gately point and $\tau$-value for the given characteristic function. We have proposed an algorithm, Density-Restricted Agglomerative Clustering (DRAC), and have provided a qualitative comparison with the existing algorithms along with its strengths and limitations.

As a future work, it would be interesting to test our method using Evolutionary game theory and Bargaining concepts. It would be worthwhile developing a characterization of games for which various game theoretic solution concepts coincide.

\section{Acknowledgement}
\label{sec:ack}
This work is an extension of a project as part of \textit{Game Theory} course. We thank Dr. Y. Narahari, the course instructor, for helping us strengthen our concepts in the subject and for guiding us throughout the making of this paper. We thank Avishek Chatterjee for mentoring our project, helping us get started with cooperative game theory and for the useful and essential criticism which helped us improve our algorithm.

\end{document}